\documentclass[twocolumn,aps,pra,superscriptaddress]{revtex4-2}

\usepackage[T1]{fontenc}
\usepackage{amsmath,amssymb,amsfonts}
\usepackage{hyperref}
\usepackage{graphicx}
\usepackage[dvipsnames]{xcolor}
\usepackage{physics}
\usepackage{tikz}
\usetikzlibrary{quantikz2}
\usepackage{bm}

\newcommand{\cA}{\mathcal{A}}

\newcommand{\cD}{\mathcal{D}}
\newcommand{\cE}{\mathcal{E}}

\newcommand{\cK}{\mathcal{K}}

\newcommand{\cN}{\mathcal{N}}

\newcommand{\cS}{\mathcal{S}}

\newcommand{\upe}{\mathrm{e}}
\newcommand{\upi}{\mathrm{i}}

\newcommand{\crot}{\textsc{crot}}

\begin{document}

\title{Single-tone drive-enhanced \crot~gate for bosonic quantum error correction}

\author{Shushen Qin}
\email{shushen.qin@u.nus.edu}
\affiliation{Centre for Quantum Technologies, National University of Singapore{, Singapore}}
\author{Long D. H. My}
\affiliation{Centre for Quantum Technologies, National University of Singapore{, Singapore}}
\author{Adrian Copetudo}
\affiliation{Centre for Quantum Technologies, National University of Singapore{, Singapore}}
\author{Amon M. Kasper}
\affiliation{Centre for Quantum Technologies, National University of Singapore{, Singapore}}
\author{Yvonne Y. Gao}
\affiliation{Centre for Quantum Technologies, National University of Singapore{, Singapore}}
\affiliation{Department of Physics, National University of Singapore, Singapore}
\author{Hui Khoon Ng}
\email{huikhoon.ng@nus.edu.sg}
\affiliation{Centre for Quantum Technologies, National University of Singapore{, Singapore}}
\affiliation{Department of Physics, National University of Singapore, Singapore}

\begin{abstract}
Bosonic error correction provides a hardware-efficient route to protected qubits for accurate quantum information processing. A critical component for error correction of rotation-symmetric bosonic (RSB) codes, like the well-known cat codes, is the two-mode controlled-rotation (\crot)~operation. The \crot~gate is useful because of its code-agnostic generality and its error-propagation properties. While the \crot~gate can in principle be composed from existing bosonic physical primitives, the desirable properties are typically lost when composed as a sequence of imperfect primitive operations. Past work has focused on direct implementations that rely on features of specific codes. Here, we present a direct route to \crot~within the circuit-QED architecture that retains its code-agnostic and error-propagation characteristics. By driving a transmon simultaneously coupled to two microwave cavities, we controllably enhance the effective nonlinearities between the microwave cavities and thus engineer the necessary two-mode interaction underlying the \crot~gate. Using only a single drive frequency, we can achieve an on-off ratio sufficient for gate implementation, while minimizing mode distortions. We provide simple analytical formulas that permit the identification of potential working regimes, further refined by exact numerics, and illustrate the efficacy of our \crot~approach in an error-correction example involving different RSB codes within the same circuit. Our \crot~gate adds to the arsenal of direct two-mode gates available for general bosonic information processing.
\end{abstract}

\maketitle

\section*{Introduction}

Quantum error correction is critical for utility-scale quantum computers. It enables active removal of errors for reliable computation, by encoding a logical qubit in a quantum state space of higher dimension, typically an ensemble of physical qubits \cite{Shor1996,Steane1996,Gottesman1998,Knill2005,Terhal2015}. Qubit-based codes, such as the currently popular surface codes, while demonstrating below-threshold operation in recent experiments \cite{Google2025,He2025,Lacroix2025}, incur significant resource overheads \cite{Fowler2012,Chamberland2017,Beverland2022,Cain2026}. In contrast, bosonic codes \cite{Cochrane1999,Gottesman2001,Mirrahimi2014,Michael2016} encode a protected logical qubit into the infinite-dimensional space of a single oscillator mode while requiring only single-mode controls per logical qubit, allowing for hardware-efficient quantum computers \cite{Leghtas2013,Terhal2020,Joshi2021,Cai2021}. Several bosonic codes have shown promising experimental performance, reaching break-even operation such that the logical qubit survives beyond the lifetime of an unprotected physical qubit in the same system \cite{Ofek2016,Hu2019,Ni2023,Sivak2023}. 

Much of the bosonic codes discussion revolves around actively correcting loss errors dominant in the physical system, and then concatenating with an outer code to correct residual phase errors \cite{Fukui2018,Vuillot2019,Noh2022,Babla2025}. In the same thread are techniques to exploit the noise structure of the two-component cat qubit \cite{Mirrahimi2014,Leghtas2015,Grimm2020,Reglade2024} to engineer a stronger noise bias, and thus delay the need for the outer code. The latter approach benefits from relaxed thresholds of biased-noise qubit codes \cite{Guillaud2019,Darmawan2021,Chamberland2022}, and permits low-overhead architectures such as concatenation with a simple repetition-code for below-threshold operation \cite{Putterman2025}. 

These efforts target dominant loss errors, but bosonic codes are in fact capable of correcting arbitrary errors. The two-component cat qubit belongs to the family of rotation-symmetric bosonic (RSB) codes \cite{Grimsmo2020}, whose number–phase structure supports the correction of arbitrary single-mode errors. Teleportation-based error-correction (EC) circuits \cite{Grimsmo2020,Hillmann2022} have been proposed for RSB codes to not only correct inherent loss errors but also phase errors due to back-action from coupling to an ancillary transmon for control \cite{Ofek2016, Rosenblum2018FTdetection}. The key two-mode gate needed for the EC circuit is the \crot~gate, $\crot(\varphi)\equiv \upe^{\upi\varphi\widehat{n}\otimes\widehat{n}}$, where $\widehat n$ is the number operator on each bosonic mode. The gate is transparent to phase errors on the data bosonic mode, and propagates loss errors to a phase on an ancillary mode. This allows both phase and loss errors to be detected, and the logical information is recovered on a third mode via teleportation. The teleportation EC circuit ensures that the output is always in the logical subspace, and thus acts also as a leakage reduction unit \cite{Aliferis2007}. The code-agnostic nature of the \crot~gate also enables an optimized code choice on the ancillary mode to maximize its robustness, improving EC performance under circuit-level noise \cite{My2025}. These various features motivate the search for a simple implementation of the \crot~gate, adding also to the arsenal of two-mode physical gates available for bosonic information processing.

Here, we focus on bosonic circuit quantum electrodynamics (cQED) \cite{Blais2021} systems, the platform behind the majority of experimental demonstrations of RSB codes. These superconducting systems benefit from the longevity of the microwave cavities, with reliable universal control of the cavity mode through dispersive coupling to a transmon \cite{Krastanov2015,Heeres2015,Heeres2017,Eickbusch2022}. Parametric beamsplitter interactions \cite{Gao2018,Gao2019,Basilewitsch2022,Chapman2023} enable universal multimode control, placing the \crot~gate within existing toolboxes as a composition of more primitive gates. Such an approach, however, is indirect and does not, in general, preserve the \crot~gate's desirable error-propagation properties. Previous implementations of two-mode gates between error-correctable bosonic modes are code dependent \cite{Chou2018,Rosenblum2018CNOT,XuY2020}, forfeiting the code-agnostic generality of the EC circuit. Existing proposals for the \crot~gate, in effect a cross-Kerr interaction between the two cavity modes, are difficult to implement: One proposed scheme relies on large cavity or transmon frequency tunability of $>\!\!\!1$GHz to turn the gate on or off \cite{ZhangY2017}, while another suggests photon-number-dependent Hamiltonian engineering \cite{Wang2021} by individually addressing each two-cavity Fock state, thus requiring a number of control frequencies that scales with the size of the logical state. Our own recent experiment \cite{Copetudo2026} successfully demonstrated the desired cross-Kerr interaction between Fock states $\ket 0$ and $\ket 2$ and their superposition through a parametric off-resonant transmon drive. However, to extend the \crot~interaction to a fully encoded RSB state, one must carefully control the higher-order distortions that can become amplified alongside the desired cross-Kerr term.

In this work, we utilize the region of sharp divergences in cavity nonlinearities, induced by a driven mediating transmon \cite{Zhang2022}, to controllably amplify the desired cross-Kerr interaction for a direct \crot~gate in a way which minimizes mode distortions, while actively suppressing self-Kerr effects. The scheme can be implemented with only transmon drives at a single frequency, without requiring other tunable physical components. Below, we present the physical system of interest and explain how cross-Kerr enhancement emerges from a transmon drive. We then show how a good working point for the \crot~gate can be identified via simple analytical formulas, and further refined by exact numerics. Finally, we demonstrate the efficacy of our code-agnostic \crot~approach in an error correction example employing different RSB codes. The Methods section provides the technical details for reproducing our results.

\section*{Results}

\subsection*{Two cavities with a common nonlinear ancilla}

\begin{figure}
\includegraphics[trim=85mm 65mm 85mm 55mm, clip, width=0.65\linewidth]{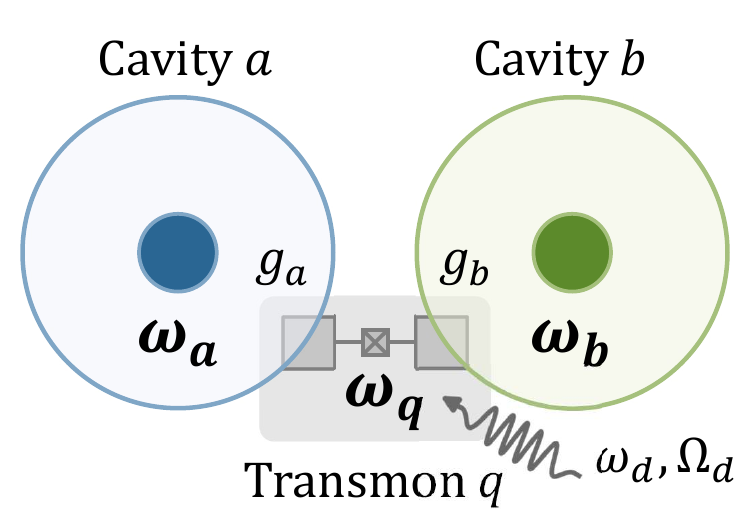}
\caption{\label{fig:sys} Circuit-QED system comprising two superconducting microwave linear cavity modes $a$ and $b$, with frequencies $\omega_a$ and $\omega_b$ respectively, coupled to a common central transmon $q$ of frequency $\omega_{q}$.}
\end{figure}

The physical system of interest comprises two linear cavity modes, $a$ and $b$ (frequencies $\omega_a$ and $\omega_b$, respectively), coupled to a common nonlinear (transmon) ancillary mode $q$ (frequency $\omega_q$) with anharmonicity $-\alpha$, where $\alpha>0$; see Fig.~\ref{fig:sys} for the typical realization in a superconducting setting. We define $\omega_{10}\equiv \omega_q-\alpha$, the frequency difference between the two lowest states of the bare (i.e., unperturbed by cavity coupling), undriven transmon. The transmon drive is at frequency $\omega_d$, detuned from $\omega_{10}$, and strength $\Omega_d$. The cavity-transmon couplings $g_a$ and $g_b$, assumed weak compared with the detunings $\delta_{a(b)}\equiv \omega_{a(b)}-\omega_{10}$, hybridize the bare cavity and transmon modes, and the dressed cavity modes inherit effective nonlinearities which can be modified via the transmon drive. 

Reference~\cite{Zhang2022} earlier described the emergence of these cavity nonlinearities using a perturbative treatment. The resulting effective self-Kerr ($K_a$, $K_b$) and cross-Kerr ($K_{ab}$) coefficients of the cavity modes, assuming the transmon remains in its ground state, take the following forms,
\begin{align}\label{eq:KerrStruct}
K_{a(b)}&= |g_{a(b)}|^4\sum \frac{q^4}{D_{a(b)}^3},\\
K_{ab}&= |g_a|^2|g_b|^2\sum \frac{q^4}{D_{ab}^3}.\nonumber
\end{align}
The detailed formulas are in the Methods section; the overall structure suffices for our discussion here. The sums range over intermediate transmon states (labeled with quantum numbers $m$), and $q^4$ denotes a product of four matrix elements, $q^{-(+)}_{m'm}\equiv \bra{\psi_{m'}}\widehat q^{(\dagger)} \ket{\psi_m}$, where $\widehat q^{(\dagger)}$ is the annihilation(creation) operator for the transmon mode $q$, and $\ket{\psi_m}$ is the bare but driven transmon eigenstate in the rotating frame of the drive at $\omega_d$, with quantum number $m$ ($m=0$ is the ground state) and eigenenergy $\epsilon_m$. That the Kerr coefficients contain a product of four $q$s, along with four powers of $g_a$ and/or $g_b$, is a consequence of their emergence from the fourth-order perturbation theory with small parameters $g_a/\delta_a$ and $g_b/\delta_b$ \cite{Zhang2022}. $D_{a(b)}^3$ denotes a product of three factors, each of the form $(\tau\delta_{a(b)d}+\epsilon_{0m})$, with $\tau=0,\pm1,\pm2$, $\delta_{a(b)d}\equiv \omega_{a(b)}-\omega_d$, and $\epsilon_{0m}\equiv \epsilon_0-\epsilon_{m}$; $D_{ab}^3$ also denotes a product of three factors, but in this case, the factors can be $(\tau\delta_{a(b)d}+\epsilon_{0m})$, with $\tau=0,\pm1$, or $(\pm \delta_{ad}\pm\delta_{bd}+\epsilon_{0m})$.

\subsection*{Drive-enhanced cross-Kerr effect}
In the absence of a transmon drive ($\Omega_d=0$), the Kerr nonlinearities are fixed and determined by the eigenstructure of the bare transmon mode. In typical bosonic information processing settings, these nonlinearities produce unwanted distortions to the computational cavity-mode states. The physical parameters are hence usually chosen to minimize the Kerr terms. When the transmon drive is turned on, the Kerr nonlinearities are affected by the changes in the eigenspectrum of the transmon mode. Specifically, in the perturbative expressions for Kerr nonlinearities, the drive changes the values of the $q$s (via modified transmon eigenstates) and the $\epsilon_m$s. This allows for active control of the Kerr effects. Reference~\cite{Zhang2022} used such a drive to suppress harmful self-Kerr nonlinearities inherent in the chosen operating physical parameters. Here, we use this active control to implement an on-demand \crot~gate on the bosonic modes: We turn on a transmon drive to amplify the cross-Kerr coefficient whenever we want to implement the gate. 

Successful gate operation has two crucial requirements: (1) a large on-off ratio, which demands a large amplification of the cross-Kerr coefficient with the drive compared to without drive; (2) the drive used to amplify the cross-Kerr must not also amplify the self-Kerr, nor the higher-order nonlinearities, as these cause mode distortions during gate operation. Here, we discuss how these two requirements can be satisfied with a careful choice of drive parameters.

We first look at the drive amplification of the cross-Kerr coefficient. In the absence of the transmon drive, the physical parameters are assumed to have been chosen such that the Kerr nonlinearities are all small, so that state distortions are negligible in standard few-microsecond protocols. This corresponds to the situation where none of the $D$ factors in the self- and cross-Kerr formulas of Eq.~\eqref{eq:KerrStruct} are small. This off-resonant scenario is the normal operating point. To amplify the cross-Kerr effect, we turn on the transmon drive to move the bare transmon eigenenergies $\epsilon_m$s so as to  induce a (near-)resonance, i.e., one (or more) of the $D$ factors in the denominator of $K_{ab}$ becomes small. Possible resonances are of two types: (i) two-cavity terms of the form $\pm\delta_{ad}\pm\delta_{bd}+\epsilon_{0m}$, and (ii) single-cavity terms of the form $\tau\delta_{a(b)d}+\epsilon_{0m}$. Two-cavity terms (i) are found only in the cross-Kerr $K_{ab}$, while the single-cavity terms (ii) occur also in the self-Kerr $K_a$ and $K_b$. We target a type (i) resonance, to amplify only the cross-Kerr, and not the self-Kerr coefficients. 

We thus want a transmon drive that would induce a resonance, $D^\textrm{res}\equiv \overline\sigma_a\delta_{ad}+\overline\sigma_b\delta_{bd}+\epsilon_{0\overline m}\simeq 0$, for some chosen $\overline\sigma_{a},\overline\sigma_{b}=\pm1$, and transmon quantum number $\overline m$. One could numerically exactly solve for $\epsilon_m$s for given transmon drive, and find a suitable resonant point among the different $\epsilon_m$ values. It is useful, however, to have analytical approximations to build intuition and guide the search for potential working points within the complex landscape. We note here already that the resonance must not be so large (i.e., $D^\textrm{res}$ not so close to 0) that one violates the perturbative regime behind the self- and cross-Kerr formulas above. This gives constraints on the allowed parameters, and further constraints arise from ensuring higher-order distortions remain small.

Near the resonance $D^\textrm{res}\simeq 0$, the cross-Kerr expression is dominated by the resonant piece in $K_{ab}$,
\begin{align}\label{eq:KABres}
K_{ab}^{\mathrm{res}}&\equiv \frac{|g_a|^{2}\,|g_b|^{2}}{D^\textrm{res}}|L|^2,\\
\textrm{with}\quad L &\equiv \sum_{m'}{\left(\frac{q^{\overline\sigma_b}_{\overline m,m'}q^{\overline\sigma_a}_{m'\!,0}\,}
        {\overline\sigma_a\delta_{ad}+\epsilon_{0m'}}+[a\leftrightarrow b]\right)}.\nonumber
\end{align}
The self-Kerr coefficients are small at this point, since none of the $D$ factors there are small (assuming no accidental resonances). We must also check higher-order distortions from the next-(6th-)order perturbative correction. The dominant term turns out to be a doubly-resonant one (see Methods), 
\begin{align}\label{eq:Kdis}
K_\textrm{dis}^\textrm{res} \!&\equiv\!\frac{|g_a|^{2}\,|g_b|^{2}}{(D^\textrm{res})^2}{\left(N_a\!+\!\tfrac{1+\overline\sigma_a}{2}\right)}\!{\left(N_b\!+\!\tfrac{1+\overline\sigma_b}{2}\right)}|L|^{2} M,\\
M&\equiv
\!\!\!\!\!\sum_{\sigma=\pm;m'}\!\!\!{\left(
   \frac{|g_a|^{2}|q^{\sigma}_{m',\overline m}|^{2}{\left(N_a\!+\!\overline\sigma_a\!+\!\tfrac{1+\sigma}{2}\right)}}{(\overline\sigma_a+\sigma)\delta_{ad}+\overline\sigma_b\delta_{bd}+\epsilon_{0m'}}+[a\!\leftrightarrow \!b]\!\right)}\nonumber.
\end{align}
Here, $N_a$ and $N_b$ are the photon numbers for the cavity modes $a$ and $b$, respectively; the comparable Kerr terms are $K_{a(b)}N_{a(b)}^2$ and $K_{ab}N_aN_b$. The double resonance $(D^\textrm{res})^2$ in $K_\textrm{dis}^\textrm{res}$ is kept in check by an additional power of $|g_{a(b)}|^2/D^\textrm{res}$ compared with $K_{ab}^\textrm{res}$, kept small for the perturbative analysis to work. This puts a restriction on how small $D^\textrm{res}$ is allowed to be. There are also singly-resonant terms at the 6th order but those are subdominant compared with $K_\textrm{dis}^\textrm{res}$.

\subsection*{Analytical working point}
Equations \eqref{eq:KABres} and \eqref{eq:Kdis} can be used to understand the parameter regimes necessary for a large cross-Kerr effect, while $K_\textrm{dis}^\textrm{res}$ remains controlled. For instance, a perturbative treatment to derive the transmon eigenstates and eigenenergies for a weak drive reveals no useful amplification (see Methods); instead, an intermediate to strong drive is needed. 

For that, we need an analytical approach for the bare-transmon eigenstructure for arbitrary drive strengths. Details are given in Methods; here, we provide only a sketch of the steps. We begin with the transmon-only Hamiltonian in the rotating-wave approximation, $H_q^\textrm{RWA}$, putting it first in normal ordering, before displacing it by a complex parameter $\beta\equiv |\beta|\upe^{\upi\phi}$: $H_\beta\equiv \cD(\beta)^\dagger H_q^\textrm{RWA}\cD(\beta)$, with $\cD(\beta)\equiv \upe^{\beta \widehat q^\dagger -\beta^*\widehat q}$ as the usual displacement operator. We choose $\beta$ such that the terms linear in $\widehat q$ and $\widehat q^\dagger$ vanish, namely, when $|\beta|$ satisfies, for $\Omega_d\equiv |\Omega_d|\upe^{\upi\phi_d}$,
\begin{equation}\label{eq:xi}
\xi(1+\xi)^2=\cK,
\end{equation}
where $\xi\equiv \alpha|\beta|^2/\delta_d$, $\delta_d\equiv \omega_d-\omega_{10}$, and $\cK\equiv \alpha|\Omega_d|^2/\delta_d^3$, and phase $\phi$ satisfies $\textrm{sgn}[\delta_d(1+\xi)]\upe^{\upi\phi}=\upe^{\upi\phi_d}$. A further Bogoliubov transformation with squeezing parameter $\zeta\equiv r\upe^{\upi\vartheta}$ is carried out, with $\zeta$ chosen so that $H_\beta$ takes a harmonic form of renormalized frequency $\delta_\beta\equiv\frac{-\delta_d(1+2\xi)}{\cosh 2r}$, with additional cubic and quartic (in $\widehat q$s) terms gathered as $V_\beta$; specifically, we require $\tanh 2r = \xi/(1+2\xi)$ and $\vartheta=2\phi$. The harmonic form is exactly solvable, and we then treat $V_\beta$ as a perturbation to obtain analytical expressions of the bare transmon eigenstructure, accurate for arbitrary drive strengths.

As a concrete example, we specialize to a particular $D^\textrm{res}$ configuration: $\overline m=1$, $\overline\sigma_a=+1$ and $\overline\sigma_b=-1$. Since $\overline\sigma_a=-\overline\sigma_b$, the drive parameters enter only $\epsilon_{01}$ in $D^\textrm{res}$: The drive frequency $\omega_d$ drops out from the linear-cavity detunings $\overline\sigma_a\delta_{ad}+\overline\sigma_b\delta_{bd}=\omega_a-\omega_b$. This resonance $D^\textrm{res}=\omega_a-(\omega_b+\epsilon_{10})\simeq 0$ corresponds to a frequency conversion between a cavity-$a$ photon and a cavity-$b$ photon, with an additional transmon excitation.

For this resonant situation, considering only the dominant $q_{mm'}^\pm$s with $|m-m'|=0,1$, the cross-Kerr coefficient can be approximated by a compact expression,
\begin{align}\label{eq:KABamp}
K_{ab}&\simeq K_{ab}^\textrm{res}=\frac{|g_a|^2|g_b|^2}{D^\textrm{res}}|L_0|^2,\\
\textrm{with}\qquad L_0&\equiv L\vert_{D^\textrm{res}=0}=\tfrac{\alpha\beta G}{\delta_\beta}{\left[-\tfrac{s}{\delta_{ad}}+\tfrac{c}{\delta_{bd}}\right]},\nonumber
\end{align}
with $G\equiv \frac{\delta_\beta}{\alpha\beta^*}(q_{1,1}^+-q_{0,0}^+)=2(c+s)^2(1+cs+2s^2)$, accurate to linear order in $|\alpha/\delta_\beta|$; here, $c\equiv \cosh r$ and $s\equiv \sinh r$. One can compare this with the no-drive situation to understand the on-off ratio for the cross-Kerr amplification; without drive, the cross-Kerr coefficient is \cite{Zhang2022} $K_{ab}^{\Omega_d=0}=-\frac{2\alpha|g_a|^2|g_b|^2(\delta_a+\delta_b)}{\delta_a^2\delta_b^2(\delta_a+\delta_b+\alpha)}+O(|g/\delta|^6)$, with $\delta_{a(b)}\equiv \omega_{a(b)}-\omega_{10}$. One can also check that none of the denominators in $M$, which contribute to higher-order distortions, are close to resonant---the denominators are all $O(\delta_{ad},\delta_{bd})$, large compared to the cross-Kerr term. 

The approximate formula [Eq.~\eqref{eq:KABamp}] for the amplified cross-Kerr coefficient gives an analytical guide to identifying a potential working point. However, because of the perturbative and resonant approximations made to obtain simple analytical formulas, the precise working point has to be optimized using exact numerics in the vicinity of the chosen region. Comparing with exact numerics for the chosen physical parameters (see below), we find $\sim$30\% discrepancy in the working point location; the discrepancies between the predicted and actual Kerr values can be larger, taking into account the contributions from the $|m-m'|>1$ values neglected in our simple formulas above. Our analytics are thus useful in identifying the region for potential operation, but exact numerics are necessary for optimal performance.

As an example, used also in the error correction discussion below, we choose typical values for the physical parameters (see our related experimental work \cite{Copetudo2026}): $\omega_a/2\pi = 4.0$GHz, $\omega_b/2\pi=4.3$GHz, $\omega_{10}(\equiv\omega_q-\alpha)/2\pi=4.65$GHz, $\alpha/2\pi=0.2$GHz, $|g_a/\delta_a|=0.04$, and $|g_b/\delta_b|=0.03$. The $g$ ratios are kept small for the perturbative description to work well, but they also enter the Kerr coefficients, and hence should not be too small. We choose $|g_a/\delta_a|\neq|g_b/\delta_b|$ to break the symmetry between cavities $a$ and $b$, but having $|g_a/\delta_a|>|g_b/\delta_b|$ is also useful in the error correction context.
From these, we want $\Omega_d$ and $\delta_d$ for good cross-Kerr amplification, while simultaneously maintaining a good on-off ratio and small higher-order distortions. One can show that neither weak nor strong drives give good cross-Kerr amplification (see Methods); we thus target $|\Omega_d/\delta_d|\sim 1$.

Keeping $\overline m=1$ and $\overline \sigma_a=-\overline\sigma_b$, the exact resonance $D^\textrm{res}=0$ demands $|\epsilon_{01}|=|\omega_a-\omega_b|=0.3(2\pi)$GHz. Thus, we have the following relations from our analytical formulas: $|\epsilon_{01}|=0.3(2\pi)$GHz, with $\epsilon_{01}$ equal to $-\delta_\beta$ plus a correction from $V_\beta$ (see Methods) and hence a function of $\delta_d$ and $|\beta|$; Eq.~\eqref{eq:xi} relates $\delta_d$, $|\beta|$, and $|\Omega_d|$; we target $|\Omega_d/\delta_d|\sim 1$. Examining these relations for the above parameter values, one can identify a potential working point: $\delta_d/\alpha=0.68$ and $|\Omega_d/\delta_d|=1$. At this point, there is significant cross-Kerr amplification, while the distortion (predicted by $M$) is only a few percent. The actual cross-Kerr value, of course, depends on how close we operate to the exact resonance point (where perturbation theory fails), to be resolved by an exact numerical optimization in the vicinity of this chosen working point. 

One could in fact reduce $M/|L_0|^2$ further by going to larger $|\Omega_d/\delta_d|$ values, but for which $\alpha/\delta_d$ grows. The perturbative treatment with $V_\beta$ small, however, requires $\alpha/\delta_d$ small, so it is not advisable to wander too far from that point, for our formulas to remain valid. One can estimate, for our chosen working point, the size of $V_\beta$ relative to the harmonic part of $H_\beta$, which turns out to be determined by the ratios (see Methods) $|\alpha/\delta_\beta|\sim 0.7$ and $|\alpha\beta/\delta_\beta|\sim 0.4$. Neither is especially small, indicating again that our perturbative approach is adequate only to identify a rough working point, and numerical optimization is crucial.

\subsection*{Numerical optimizations}

\begin{figure*}
\includegraphics[width=\textwidth]{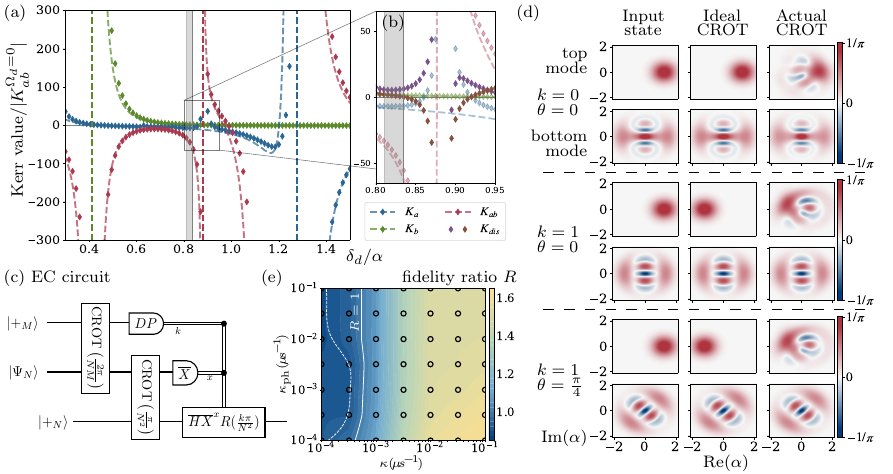}
\caption{\label{fig:Numerics} (a) Kerr spectrum (in units of the no-drive cross-Kerr) in the weak-coupling regime at fixed effective amplitude $\left|\frac{\Omega_d}{\delta_d}\right|=1$. Dispersive structures arise from drive-induced resonances. The dashed green and blue vertical lines mark single-cavity resonances where both cross- and (unwanted) self-Kerr effects are amplified. The desired working point lies near the two-cavity resonance marked by the dashed magenta vertical line. (b) The immediate vicinity of the chosen resonance, overlaid with the numerical sixth-order terms $K_\textrm{dis}$ $\propto N_a^2 N_b$ (purple) and $N_a N_b^2$ (brown) that lead to distortions. The gray shaded region identifies a range of drive detuning for which the higher-order distortions are mild, while the cross-Kerr coefficient already experiences significant growth close to the resonance. Our chosen \crot~working point is at $\delta_d/\alpha=0.835$ (gray vertical line). (c) Teleportation-based error-correction (EC) circuit for rotation-symmetric bosonic (RSB) codes, employing two \crot~gates [\crot$(2\pi/NM)$ and \crot$(\pi/N^2)$, with $M$ and $N$ as the RSB code orders], two measurements (DP $=$ discrete-phase measurement; $\overline X=$ logical-$X$ measurement), and a final classically controlled single-mode gate to complete the teleportation. (Figure adapted from Ref.~\cite{Grimsmo2020}.) (d) Wigner diagrams of input and output [after \crot$(2\pi/NM)$] states at our chosen working point, for different loss error $k$ and phase error $\theta$ values. (e) Memory task: Ratio $R$ of the average fidelities, after EC with actual \crot~gates versus the unencoded physical qubit, for different loss and dephasing rates $\kappa$ and $\kappa_{\text{ph}}$. The small black circles are numerically simulated data points; the contour plot interpolates across those points. The solid white line marks $R=1$ for EC with actual \crot~gates; the dashed white line gives the analogous line for EC with ideal \crot s.
}
\end{figure*}

The exact numerics involves first a full diagonalization of the two-cavity--ancilla Hamiltonian, extracting the exact dressed eigenenergies and eigenstates. Focusing only on the dressed states adiabatically connected to the undriven transmon ground state ($m=0$), we use the numerically obtained energy differences to construct a system of linear equations to solve for the linear (Stark) shifts and the nonlinear Kerr coefficients. 

Figure \ref{fig:Numerics} shows the numerically obtained Kerr coefficients (markers), superimposed on the results from perturbation theory (dashed lines, using the full expressions for Eq.~\eqref{eq:KerrStruct} and exact numerical calculation of the driven bare transmon eigenstructure), for our chosen physical parameters with $|\Omega_d/\delta_d|=1$, for the region around the analytical $\delta_d/\alpha=0.68$ value. We include a larger range of $\delta_d/\alpha$ values to show the broad accuracy of the perturbative approach: The perturbative description shows good predictability of the Kerr coefficients, with deviations, as to be expected, near resonances. The resonance identified using the exact transmon eigenstates falls near the dashed magenta vertical line, at $\delta_d/\alpha\simeq 0.88$, compared with our analytical approximation of 0.68, with discrepancies coming from considering only the resonant terms and the perturbative treatment for the transmon-only eigenstructure. 
The solid gray vertical line marks the final chosen working point, with $\delta_d/\alpha =0.835$, just to the left of the exact resonance. At this point, the amplification of the cross-Kerr coefficient is sufficient for gate implementation with an on-off ratio exceeding $60$, while keeping the higher-order distortions under control; see zoomed-in plot Fig.~\ref{fig:Numerics}(b).

In Fig.~\ref{fig:Numerics}(b), one can observe a small dispersive feature in the cavity-$a$ self-Kerr value $K_a$ (blue markers) near that two-cavity resonance, coming from the 6th-order correction. While it is an order lower in magnitude compared with the cross-Kerr value due to the suppression by small $g/\delta$, any nonzero self-Kerr effect causes mode distortions. Following Ref.~\cite{Zhang2022}, that self-Kerr can be actively canceled by adding a driven side transmon $q_a$ coupled only to cavity $a$. For our chosen working point, it suffices to have a side transmon of frequency $\omega_{q_a,10}=\omega_{10}+1.5\alpha$, coupled to cavity $a$ with strength $g_a^{q_a}=0.01(\omega_a-\omega_{q_a,10})$, and driven at the same frequency as the central transmon $q$ at an amplitude $|\Omega_d^{q_a}/\delta_d^{q_a}|=\sqrt{0.6}$. This cancels the excess self-Kerr induced by the drive on the central transmon. A driven side transmon can similarly be added for cavity $b$ if desired, though at our working point, the induced cavity-$b$ self-Kerr is already quite small.

\subsection*{Error correction for bosonic codes}
As a demonstration of our cross-Kerr gate, we explore its use as a \crot~gate in the teleportation-based EC circuit of Refs.~\cite{Grimsmo2020, Hillmann2022, My2025} for RSB codes, which include the well-known cat and binomial codes. The \crot-based teleportation circuit applies across different RSB codes, and this showcases the code-agnostic flexibility of our cross-Kerr implementation. 

The EC circuit of interest is the ``hybrid'' circuit introduced in Ref.~\cite{Grimsmo2020}, capable of simultaneously detecting loss and phase (rotation in phase space) errors; see Fig.~\ref{fig:Numerics}(c). Three modes are involved, each encoded in a (possibly different) RSB code of some code order: The top rail carries an ancillary mode with code order $M$; the middle and bottom rails carry the input and output data modes, respectively, both assumed to have the same code order $N$. In the first step, a \crot~gate propagates $k$ photon losses ($\widehat a^k$) on the input data mode to a $-2\pi k/NM$ phase rotation on the ancillary mode, which is subsequently measured (using a discrete phase (DP) measurement; see Refs.~\cite{Grimsmo2020,Hillmann2022}) to determine the value of $k$. The second \crot~begins a one-bit teleportation, transferring the encoded information to the output data mode while leaving behind any phase error on the input data rail which enters a logical $\overline{X}$ measurement. A final measurement-outcome-controlled gate on the output rail completes the teleportation. 

The set of correctable errors is of the form $\{R(\theta)\widehat{a}^k\}$, with $k<N$ and continuous phase rotation $R(\theta)$ with $\theta$ in a chosen $\frac{\pi}{N}$ range. The ancillary mode is set to have $M=1$ for maximal robustness against phase errors in the DP measurement \cite{My2025}. For our illustration below, we choose the input and output data modes to be in the $N=2$ binomial code, while the ancillary mode is an $M=1$ cat code, for which the initial ancilla state is close to a simple coherent state.

For the first \crot~gate, we set the drive parameters to the chosen working point from the earlier discussion, for a gate time $T_{\crot 1}=\frac{\pi}{K_{ab}}$. This implements a $\crot\left(\frac{2\pi}{NM}\right)$ between a $\ket{+_{M=1,\textrm{cat}}}\sim\ket{\alpha_\textrm{coh}}$ in the top ancillary mode (plays the role of mode $a$ in our earlier discussion) and $\ket{+_{N=2,\textrm{bin}}}\sim \ket{0}+\sqrt{2}\ket{2}+\ket{4}$ binomial-code state in the input data mode (mode $b$ in our earlier discussion). The columns in Fig.~\ref{fig:Numerics}(d) show the states before and after the \crot~gate, comparing the results from our cross-Kerr-amplified version (with the full Hamiltonian, including any higher-order distortions, and self-Kerr cancellation drives on both modes) to that from an ideal $\crot$~gate. From the Wigner plots, the data rail output from the actual \crot~matches well that from the ideal gate (with fidelities $\approx0.96, 0.99, 0.99$ \footnote{Note that the no-error state has slightly lower fidelity as it has higher weight on the higher Fock states, which are more sensitive to imperfections in the \crot~gates.} respectively, for the cases of no error, single loss error, and loss and phase errors). The ancillary mode, however, suffers significant distortions, but that is not detrimental for EC: The ancillary mode directly enters the DP measurement that only distinguishes whether the state lies on the right or left in phase space (see ideal \crot~outputs), and is thus tolerant of distortions in the mode details. In fact, the higher ancilla distortion is a choice, where a stronger drive for higher cross-Kerr---and hence a faster gate---reduces the higher-order distortions on the data mode, while the mode with the higher distortions (from the asymmetry in the cavity-transmon couplings) is designated as the ancilla. One can also observe in the last row of Fig.~\ref{fig:Numerics}(d) that a rotation error of $\upe^{\upi\frac{\pi}{4}}$ passes through the \crot~gate without propagating to the ancilla, a property key to the correct functioning of \crot.

The second \crot~gate is implemented in a similar manner, though it is now between two modes encoded in the same code order. We lower the drive amplitude slightly to reduce distortion in both modes, but the drive configuration is otherwise kept identical to the first \crot~so that a single drive frequency suffices for the entire EC circuit, for the \crot~gates and the self-Kerr cancellation. 

To evaluate the performance of the EC circuit, we numerically simulate the memory task, where a quantum state is left idle for a duration $T_\textrm{wait}$ subjected to noise, and then one round of EC is carried out. We compute the fidelity between the corrected output and the input, averaged over the cardinal logical states, $F=\frac{1}{6}\sum_{\ell=\pm X, \pm Y, \pm Z} {}_L\bra{\ell}\textrm{EC}\circ\cN(\ket{\ell}_{\!L}\!\!\bra{\ell})\ket{\ell}_L$, where $\cN(\cdot)$ is the noise and $\textrm{EC}(\cdot)$ denotes the EC circuit implemented using our $\crot$~gates. We consider correctable errors on the input data mode, i.e. $k<N, -\pi/2N<\theta<\pi/2N$, and the EC circuit is noiseless except for imperfections from the \crot s. The results are benchmarked against the average fidelity $F_\textrm{ideal}$ for ideal \crot~gates, as well as the fidelity $F_\textrm{bare}$ for a bare, unencoded qubit (in the subspace spanned by the Fock states $\ket{0}$ and $\ket{1}$). This \crot-distortion-only perspective complements earlier studies that considered an ideal EC circuit for code-capacity performance \cite{Grimsmo2020}, the effects of imperfect measurements \cite{Hillmann2022}, and general full circuit-level-noise fault tolerance \cite{My2025}. 

Figure \ref{fig:Numerics}(e) shows the fidelity ratio $R\equiv F/F_\textrm{bare}$ over a range of practically relevant loss and dephasing rates $\kappa$ and $\kappa_{\mathrm{ph}}$. For each data point (small black circles), the idle duration $T_\textrm{wait}$ is optimized to balance the memory $\kappa$ and $\kappa_\mathrm{ph}$ noise versus the noise from the EC circuit; this ensures we are not doing noisy EC too often when the memory noise is low while EC occurs earlier when the memory noise is high. The plot shows that the ratio $R$ exceeds $1$ across all $\kappa_{\mathrm{ph}}$ at high $\kappa$, corresponding to break-even operation for regions of high loss---a high $\kappa$ results in poor bare-qubit fidelity, while the EC circuit retains correction capabilities even if the $\crot$~gate is imperfect. The EC performance is limited for small dephasing rates, even with ideal EC, as the chosen experimentally realistic average photon number of $\overline n=2$ gives limited phase distinguishability in the code states. The gain from EC thus outweighs the inherent errors only at moderate to high dephasing. Compared with ideal $\crot$~gates, the actual $\crot$s reduce EC performance by $\sim\!\!7\%$ in regions of weak noise, and by $\sim\!\!12\%$ in regions of strong noise, thus shifting the break-even pseudo-threshold to higher noise; see the white lines in Fig.~\ref{fig:Numerics}(e).

\section*{Discussion}
The \crot~gate is a key component in error correction on rotation-symmetric bosonic codes, necessary to realize the full hardware-efficient potential of codes like cat and binomial codes. Rather than prior code-specific attempts, or relying on gate primitives from a universal set, we proposed a direct route to implementing a \crot~using only a transmon drive at a single frequency. The transmon drive induces a resonance that amplifies the effective cross-Kerr interaction between the cavity modes, generating dynamics that natively yield a \crot~gate, thus ensuring the error-propagation and code-agnostic features of the original \crot~definition. We showed how analytical formulas provide a means of identifying an appropriate regime of drive parameters for cross-Kerr amplification with controlled higher-order distortions; exact numerical optimizations in the vicinity of the analytical working point provide further improvements. Our example of RSB error correction illustrates the performance of our cross-Kerr-amplified \crot~in the teleportation EC circuit, enhanced by self-Kerr cancellation drives at the same frequency. The \crot~gate itself, of course, also adds to the variety of direct two-mode physical bosonic gates.

With only a single drive frequency, our \crot~performance is ultimately limited by the higher-order distortions that come along with the cross-Kerr amplification. We showed that it is possible to find single-drive-frequency working points where those distortions leave the error-correction performance largely intact, but the amount of amplification, and hence the final gate speed, is constrained by having to keep those higher-order effects in control---one could have operated closer to the resonance if not for the associated rapid growth in distortions; see Fig.~\ref{fig:Numerics}(b). A natural next step would be to explore active suppression of higher-order effects using another transmon drive at a different frequency. Adding another frequency to the system can lead to beating effects which will make the analysis more complex (see, for example, Ref.~\cite{Zhang2019}), though the resulting Floquet structure is much richer and can potentially provide a wider range of optimal working points. We leave that as an open avenue for future investigation.

\section*{Methods}

\noindent\textbf{Physical system}\\[0.5ex]
The Hamiltonian for the two-cavity--transmon system is $H = H_\textrm{cav} + H_q + H_\textrm{I}$, with (in units of $\hbar$)
\begin{align}\label{H_lab}
H_\textrm{cav} &=  \omega_a\widehat{a}^\dagger\widehat{a} + \omega_b\widehat{b}^\dagger\widehat{b},\\
H_q &=  \omega_{q}\widehat{q}^\dagger\widehat{q}-\tfrac{\alpha}{12}(\widehat{q}+\widehat{q}^\dagger)^4+(\Omega_d^*\upe^{\upi\omega_d t}\!-\!\Omega_d\upe^{-\upi\omega_d t})(\widehat q\!-\!\widehat q^\dagger)\nonumber\\
H_\textrm{I} &=  (g_a\widehat{q}^\dagger+ g_a^*\widehat{q})(\widehat{a}^\dagger+\widehat{a})+(g_b\widehat{q}^\dagger+ g_b^*\widehat{q})(\widehat{b}^\dagger+\widehat{b}).  \nonumber
\end{align}
Here, $\widehat{a}(\widehat{a}^\dagger)$ and $\widehat {b}(\widehat b^\dagger)$ are the annihilation(creation) operators for the two linear cavity modes; $\widehat{q}(\widehat q^\dagger)$ is for the transmon mode. $H_\textrm{cav}$ describes the bare linear cavity modes with frequencies $\omega_a$ and $\omega_b$. $H_q$ describes the transmon ancillary mode with frequency $\omega_{q}$ and anharmonicity $\alpha$; $\Omega_d$ gives the strength of the transmon drive at frequency $\omega_d$. In $H_q$, the cosine potential of the transmon has been truncated at the standard fourth-order term for a weakly anharmonic transmon. $H_\textrm{I}$ describes the dipole interaction between the cavities and the transmon, with coupling constants $g_a$ and $g_b$.

\bigskip
\noindent\textbf{Drive frame}\\[0.5ex]
We move into a frame rotating at the transmon drive frequency $\omega_d(>0)$. The total Hamiltonian becomes $U_d H U_d^\dagger - H_d$, with $H_d\equiv\omega_d(\widehat a^\dagger \widehat a+\widehat b^\dagger \widehat b+\widehat q^\dagger \widehat q)$ and $U_d\equiv \upe^{\upi H_d t}$. The rotating wave approximation (RWA) applies (for small photon numbers) when $|\delta_{\cdot d}|,|\alpha|,|g_{a(b)}|,|\Omega_d|\ll\omega_d$, where $\delta_{\cdot d}\equiv\omega_\cdot-\omega_d$ for $\cdot=a,b,$ or $q$. We neglect rapidly rotating terms to arrive at the time-independent RWA Hamiltonian,
\begin{align}
H^\textrm{RWA} &= \delta_{ad}\widehat{a}^\dagger\widehat{a} + \delta_{bd}\widehat{b}^\dagger\widehat{b} + H_q^\textrm{RWA} +H_\textrm{I}^\textrm{RWA},\nonumber\\
H_q^\textrm{RWA} &= \delta_{qd}\widehat{q}^\dagger\widehat{q}-\tfrac{\alpha}{2}(\widehat{q}^\dagger\widehat{q}+1)\widehat{q}^\dagger\widehat{q}+(\Omega_d^*\widehat q+\textrm{h.c.})\nonumber\\
H_\textrm{I}^\textrm{RWA} &= (g_a\widehat{a}+g_b\widehat{b})\widehat{q}^\dagger + (g_a^*\widehat{a}^\dagger+g_b^*\widehat{b}^\dagger)\widehat{q}.\label{eq:RWA_H}
\end{align}

\noindent\textbf{Dressed modes and effective eigenstructure}\\[0.5ex]
$H_q^\textrm{RWA}$ is time-independent with eigenstates $\ket{\psi_m}$ and energies $\epsilon_m$. At zero coupling ($g_a,g_b=0$), the eigenstates of $H^\textrm{RWA}$ are product states $\ket{\psi_m,N_a,N_b}\equiv\ket{\psi_m}\otimes\ket{N_a}\otimes\ket{N_b}$, where $\ket{N_{a(b)}}$ is the cavity Fock state with photon number $N_{a(b)}$, with energy $\cE_m^{(0)}(N_a,N_b)=N_a\delta_{ad}+N_b\delta_{bd}+\epsilon_m$.
When the coupling is turned on ($g_a,g_b\neq 0$), assuming a nondegenerate system, every zero-coupling eigenstate $\ket{\psi_m,N_a,N_b}$ is adiabatically connected to an eigenstate of $H^\textrm{RWA}$---the ``dressed'' state $\ket{\overline{\psi_m,N_a,N_b}}$. We follow Ref.~\cite{Zhang2022} and define dressed cavity modes $\widehat A$ and $\widehat B$, and write $H^\textrm{RWA} = \sum_{m} \cE_m(\widehat{N}_A,\widehat{N}_B)\widehat{P}_m$, with $\widehat P_m\equiv \sum_{N_a,N_b}\ketbra{\overline{\psi_m,N_a,N_b}}$. For clarity here, from here on, we will continue to denote the dressed modes as $a$ and $b$; we refer the reader to Ref.~\cite{Zhang2022} for a careful treatment that keeps track of the dressed modes. $\cE_m$ is the generalization of the zero-coupling $\cE_m^{(0)}$ to the $g_a,g_b\neq 0$ situation, both parameterized by the quantum numbers $N_a$ and $N_b$ for given $m$. While $\cE_m^{(0)}$ is linear in $N_a$ and $N_b$, the nonzero coupling gives rise to nonlinearities in $N_a$ and $N_b$ in $\cE_m$, dependent on the state of the ancilla.

\bigskip
\noindent\textbf{Kerr coefficients from perturbation theory}\\[0.5ex]
We work in the weak-coupling regime, where $|g_{a(b)}|\ll |\delta_{a(b)d}-\delta_{qd}| \approx |\delta_{a(b)}|$. The cavity-transmon hybridization is small and $H_\mathrm{I}^\textrm{RWA}\equiv V$ can be treated as a perturbation to the uncoupled system. We express the eigenenergies and states as $\cE_\mu =  \sum_{j=0}^\infty\cE_\mu^{(j)}$ and $\ket{\psi_\mu}\equiv\ket{\overline{\psi_m,N_a,N_b}}=\sum_{j=0}^\infty |\psi_\mu^{(j)}\rangle$, with $\mu\equiv(m, N_a, N_b)$. The $j$th term is accurate to $j$ insertions of $V$ and is a polynomial in $N_a$ and $N_b$. $\cE_\mu^{(0)}\equiv \cE_m^{(0)}(N_a,N_b)$ and $|\psi_\mu^{(0)}\rangle=\ket{\psi_m,N_a,N_b}$ are the uncoupled energy and state. Then, using standard Green's function approach to perturbation theory, we recover the expressions for $\cE_\mu^{(j\leq 4)}$ in Ref.~\cite{Zhang2022}, now written in a compact form, and extended to $j=6$ needed for our analysis here:
\begin{align}\label{eq:PT}
\cE_\mu^{(2)}&=\sum_{\mu_1\neq \mu}\frac{V_{\overline{\mu\mu_1\mu}}}{\Delta_{\mu:\mu_1}},\\
\cE_\mu^{(4)}&=\sum_{\mu_1,\mu_2,\mu_3\neq\mu}\frac{V_{\overline{\mu\mu_3\mu_2\mu_1\mu}}}{\Delta_{\mu:\mu_3\mu_2\mu_1}}-\cE_\mu^{(2)}\sum_{\mu_1\neq\mu}\frac{V_{\overline{\mu\mu_1\mu}}}{\Delta_{\mu:\mu_1^2}},\nonumber\\
\cE_\mu^{(6)}&=\!\!\!\!\sum_{\mu_1,\mu_2,\ldots,\mu_5\neq \mu}\!\!\frac{V_{\overline{\mu\mu_5\mu_4\ldots\mu_1\mu}}}{\Delta_{\mu:\mu_5\mu_4\ldots\mu_1}}-\!\!\sum_{\mu_1\neq \mu}\!\frac{V_{\overline{\mu\mu_1\mu}}}{\Delta_{\mu:\mu_1^2}}{\biggl[\cE_\mu^{(4)}\!\!-\!\frac{(\cE_\mu^{(2)})^2}{\Delta_{\mu:\mu_1}}\biggr]}
\nonumber\\
&\quad -\cE_\mu^{(2)}\!\!\!\!\sum_{\mu_1,\mu_2,\mu_3\neq \mu}\!\!\frac{V_{\overline{\mu\mu_3\mu_2\mu_1\mu}}}{\Delta_{\mu:\mu_3\mu_2\mu_1}}{\left[\frac{1}{\Delta_{\mu:\mu_1}}\!+\!\frac{1}{\Delta_{\mu:\mu_2}}\!+\!\frac{1}{\Delta_{\mu:\mu_3}}\right]},\nonumber
\end{align}
and $\cE_\mu^{(j)}=0$ for $j$ odd.
Here, $\Delta_{\mu:\mu_n^{p_n}\ldots\mu_2^{p_2}\mu_1^{p_1}}\equiv  (\Delta_{\mu:\mu_n})^{p_n}\ldots(\Delta_{\mu:\mu_2})^{p_2}(\Delta_{\mu:\mu_1})^{p_1}$ with $\Delta_{\mu:\mu'}\equiv \cE_\mu^{(0)}-\cE_{\mu'}^{(0)}$, and $V_{\overline{\mu_n\ldots\mu_3\mu_2\mu_1}}\equiv V_{\mu_n\mu_{n-1}}\ldots V_{\mu_3\mu_2} V_{\mu_2\mu_1}$, with $V_{\mu'\mu}\equiv \langle\psi_{\mu'}^{(0)}|V|\psi_{\mu}^{(0)}\rangle=V_{\mu\mu'}^*$ as the matrix elements for $V$,
\begin{align}
&\quad V_{\mu'\mu}=V_{(m',N_a',N_b')(m,N_a,N_b)}\label{eq:Vmpm}\\
&=\delta_{N_b',N_b}\!{\left[g_aq_{m'm}^+\delta_{N_a',N_a-1}\sqrt{\!N_a}\!+\!g_a^*q_{m'm}^-\delta_{N_a',N_a+1}\sqrt{\!N_a\!+\!1}\right]}\nonumber\\
&\qquad +[a\leftrightarrow b]\nonumber,
\end{align}
where $[a\leftrightarrow b]$ denotes the same term as in the previous line, but with the roles of $a$ and $b$ swapped. Here, $q^{+}_{m'm}\equiv \bra{\psi_{m'}}q^\dagger \ket{\psi_m}$ and $q^{-}_{m'm}\equiv \bra{\psi_{m'}}q \ket{\psi_m}=(q^+_{mm'})^*$.

\vspace{4pt}\break
The $\cE^{(2)}_\mu$ corrections give only terms linear in $N_a$ or $N_b$. The lowest-order nonlinear correction to the cavity energy comes from the fourth-order term $\cE_\mu^{(4)}$, with the coefficient of the $N_{a(b)}^2$ term recognized as the self-Kerr nonlinearity $K_{m;a(b)}$ of the (dressed) cavity mode $a(b)$, while the coefficient of $N_aN_b$ is the cross-Kerr nonlinearity $K_{m;ab}$. From Eq.~\eqref{eq:PT}, the self-Kerr for mode $a$, for given transmon quantum number $m$, can be straightforwardly computed to be,
\begin{align}
K_{m;a}&= |g_a|^4\Biggl\{-\chi_{m;a}^{(1)}\,\chi_{m;a}^{(2)}\label{eq:selfKerr}\\
&\hspace*{-0.5cm}+\!\!\!\sum_{\substack{m_1,m_2,m_3\\
\sigma=\pm}}\!\frac{q^{-\sigma}_{m\,m_3}\,q^{-\sigma}_{m_3m_2}\,q^{\sigma}_{m_2m_1}\,q^{\sigma}_{m_1m}}{(\sigma\delta_{ad}\!\!+\!\epsilon_{mm_1}\!)(2\sigma\delta_{ad}\!\!+\!\epsilon_{mm_2}\!)(\sigma\delta_{ad}\!\!+\!\epsilon_{mm_3}\!)}\nonumber\\
&\hspace*{-0.5cm}+\!\!\sum_{\substack{m_1,m_3\\m_2\neq m\\\sigma_1,\sigma_3=\pm}}\!\!\!\frac{q^{-\sigma_3}_{m\,m_3}\,q^{\sigma_3}_{m_3m_2}\,q^{-\sigma_1}_{m_2m_1}\,q^{\sigma_1}_{m_1m}}{(\sigma_1\delta_{ad}\!\!+\!\epsilon_{mm_1})\epsilon_{mm_2}(\sigma_3\delta_{ad}\!\!+\!\epsilon_{mm_3})}\Biggr\},\nonumber
\end{align}
where $m_1,m_2$, and $m_3$ range over the transmon quantum numbers, $\epsilon_{mm'}\equiv \epsilon_m-\epsilon_{m'}$, and $\chi_{m;a}^{(\ell)} \equiv  \sum_{m_1;\sigma=\pm}|q^{\sigma}_{m_1m}|^2/(\sigma\delta_{ad}+\epsilon_{mm_1})^\ell$.
The self-Kerr $K_{m;b}$ for mode $b$ has an identical expression, with the roles of $a$ and $b$ swapped, and $\chi_{m;b}^{(\ell)}$ is defined analogously. The cross-Kerr coefficient is

\begin{widetext}
\begin{align}
K_{m;ab} &= |g_a|^2|g_b|^2\Biggl\{-\chi_{m;a}^{(1)}\chi_{m;b}^{(2)}+\sum_{\substack{m_1,m_2,m_3\\\sigma_a,\sigma_b=\pm}}\frac{q^{-\sigma_b}_{m\,m_3}\,q^{-\sigma_a}_{m_3m_2}\,q^{\sigma_b}_{m_2m_1}\,q^{\sigma_a}_{m_1m}}{(\sigma_a\delta_{ad}+\epsilon_{mm_1})(\sigma_a\delta_{ad}+\sigma_b\delta_{bd}+\epsilon_{mm_2})(\sigma_b\delta_{bd}+\epsilon_{mm_3})}\label{eq:crossKerr}\\
&\hspace*{-1cm} +\!\!\!\!\sum_{\substack{m_1,m_3\\m_2\neq m\\\sigma_a,\sigma_b=\pm}}\!\!\!\!\frac{q^{-\sigma_b}_{m\,m_3}\,q^{\sigma_b}_{m_3m_2}\,q^{-\sigma_a}_{m_2m_1}\,q^{\sigma_a}_{m_1m}}{(\sigma_a\delta_{ad}\!\!+\!\epsilon_{mm_1})\epsilon_{mm_2}(\sigma_b\delta_{bd}\!\!+\!\epsilon_{mm_3})}
+\!\!\!\sum_{\substack{m_1,m_2,m_3\\\sigma_a,\sigma_b=\pm}}\!\!\frac{q^{-\sigma_a}_{m\,m_3}\,q^{-\sigma_b}_{m_3m_2}\,q^{\sigma_b}_{m_2m_1}\,q^{\sigma_a}_{m_1m}}{(\sigma_a\delta_{ad}+\epsilon_{mm_1})(\sigma_a\delta_{ad}+\sigma_b\delta_{bd}+\epsilon_{mm_2})(\sigma_a\delta_{ad}+\epsilon_{mm_3})}\Biggr\}+~[a\leftrightarrow b].\nonumber
\end{align}
\end{widetext}
Setting $m=0$ for the transmon ground state gives the form of the Kerr coefficients in the main text [Eq.~\eqref{eq:KerrStruct}].

\bigskip
\noindent\textbf{Drive-enhanced cross-Kerr}\\[0.5ex]
We focus on the $m=0$ case. As described in the main text, to amplify the cross-Kerr effect, we operate near the two-cavity resonance $D^\textrm{res}\equiv \overline\sigma_a\delta_{ad}+\overline\sigma_b\delta_{bd}+\epsilon_{0\overline m}\simeq 0$, for some chosen $\overline\sigma_{a},\overline\sigma_{b}=\pm1$, and transmon quantum number $\overline m$. The resonant term in $K_{0;ab}^{(2)}$ gives Eq.~\eqref{eq:KABres} in the main text, while the doubly-resonant term in $\cE^{(6)}_{(0, N_a,N_b)}$ leads to Eq.~\eqref{eq:Kdis}. To proceed analytically, we need expressions for the transmon-only eigenstructure from $H_q^\textrm{RWA}$, which gives the $q$s and $\epsilon$s in those equations.

\medskip
\noindent\underline{Weak drive}. In the weak-drive limit, we treat the drive term $V_q\equiv \Omega_d^*\widehat q+\Omega_d\widehat q^\dagger$ as a perturbation to $H_{q,0}\equiv \delta_{qd}\widehat{q}^\dagger\widehat{q}-\tfrac{\alpha}{2}(\widehat{q}^\dagger\widehat{q}+1)\widehat{q}^\dagger\widehat{q}$. This requires $|\Omega_d|\ll |\delta_{qd}|,\alpha$. The unperturbed eigenstates and eigenenergies are $|\psi_m^{(0)}\rangle=\ket m$ the Fock states, and $\epsilon_m^{(0)}=m[\delta_{qd}-\frac{\alpha}{2}(m+1)]$. Now, since $V_q$ is off-diagonal in the Fock basis, $\epsilon_m$ is modified only at second order. Recalling that $\epsilon_m$ is the knob that we tune to induce a resonance, this already points to a poor on-off ratio: The targeted resonant point goes towards zero only by $O(|\tfrac{\Omega}{\delta,\alpha}|^2)$ in the presence of the drive (the ``on'' setting), compared to without drive, and $|\tfrac{\Omega}{\delta,\alpha}|\ll 1$ in this weak-drive limit. We hence do not expect a practical \crot~working point with a weak drive.

\noindent\underline{Arbitrary drive strength}. We thus turn to an analysis that can handle arbitrary drive strengths. We begin with $H_q^\textrm{RWA}$, putting it first in normal ordering before displacing it by a complex parameter $\beta\equiv |\beta|\upe^{\upi\phi}$, to get $H_\beta\equiv \cD(\beta)^\dagger H_q^\textrm{RWA}\cD(\beta)$,
with $\cD(\beta)\equiv \upe^{\beta \widehat q^\dagger -\beta^*\widehat q}$ so that $\cD(\beta)^\dagger \widehat q\cD(\beta)=\widehat q+\beta$ as usual. The terms linear in $\widehat q$ and $\widehat q^\dagger$ are $[\delta_{qd}-\alpha(1+|\beta|^2)]\beta \widehat q^\dagger +\Omega_d\widehat q^\dagger + \textrm{h.c.}$, which we demand to vanish by choosing $|\beta|$ to satisfy, for $\Omega_d\equiv |\Omega_d|\upe^{\upi\phi_d}$ [Eq.~\eqref{eq:xi}], $\xi(1+\xi)^2=\cK$, where $\xi\equiv \alpha|\beta|^2/\delta_d$, $\delta_d\equiv \omega_d-\omega_{10}$, and $\cK\equiv \alpha|\Omega_d|^2/\delta_d^3$, and phase $\phi$ such that $\textrm{sgn}[\delta_d(1+\xi)]\upe^{\upi\phi}=\upe^{\upi\phi_d}$, as given in the main text. Then, $H_\beta\equiv -\delta_d[(1+2\xi)\widehat q^\dagger \widehat q+\tfrac{1}{2}\xi(\upe^{\upi2\phi}\widehat q^\dagger{}^2+\upe^{-\upi 2\phi}\widehat q^2)]+V_\beta$, with $V_\beta\equiv V_3 + V_4$, where $V_3\equiv -\alpha(\beta \widehat q^\dagger{}^2\widehat q+\beta^*\widehat q^\dagger \widehat q^2)$ and $V_4\equiv -\tfrac{\alpha}{2}\widehat q^\dagger{}^2\widehat q^2$ contain, respectively, the terms cubic and quartic in $\widehat q$ and $\widehat q^\dagger$, dropping irrelevant constants.

We pause to first understand the structure of Eq.~\eqref{eq:xi}. Assuming $\alpha>0$, observe that $\xi$, $\cK$, and $\delta_d$ have the same signs. At zero drive, i.e., $|\Omega_d|=0$, $\cK=0$, and $\xi=0$. When $\cK>0$ or $<-4/27$, the cubic equation has one real root; when $\cK\in[-4/27,0]$, there are multiple roots. In practice, we ramp up the drive amplitude from zero, thus accessing $\xi$ values connected to $\xi=0$. The region with multiple solutions for $\xi$ presents a potentially unstable configuration that we want to avoid in case of a fast drive ramp-up. This thus points to focusing on $\cK>0$, and correspondingly having $\delta_d>0$ (and $\xi>0$). 

Looking only at the quadratic terms, i.e., $H_{\beta,0}\equiv H_\beta-V_\beta$, we perform a squeezing (Bogoliubov) transformation by defining new modes, $\widehat Q\equiv \cS(\zeta)^\dagger \widehat q \cS(\zeta)$, such that $[\widehat Q,\widehat Q^\dagger]=1$. Here, $\cS(\zeta)\equiv \upe^{-\frac{1}{2}(\zeta \widehat q^\dagger{}^2-\zeta^*\widehat q^2)}$ is the squeezing operator, and $\zeta\equiv r\upe^{\upi\vartheta}$ is the squeezing parameter. Standard calculations give $\widehat Q=(\cosh r)\widehat q-\upe^{\upi\vartheta}(\sinh r)\widehat q^\dagger$ and $\widehat q=(\cosh r)\widehat Q+\upe^{\upi\vartheta}(\sinh r)\widehat Q^\dagger$, allowing us to express $H_{\beta,0}$ in terms of $\widehat Q$ and $\widehat Q^\dagger$. The squeezing parameter is chosen to satisfy ($r\equiv |\zeta|\geq 0$) $\tanh 2r=\frac{\xi}{1+2\xi}$ and $\vartheta=2\phi$, so that the $\widehat Q^2$ and $\widehat Q^\dagger{}^2$ terms in $H_{\beta,0}$ vanish, leaving only a pure harmonic term $H_{\beta,0}=\delta_\beta \widehat Q^\dagger \widehat Q$ at the renormalized frequency $\delta_\beta\equiv \frac{-\delta_d(1+2\xi)}{\cosh 2r}=-\delta_d\sqrt{(1+\xi)(1+3\xi)}$. 

$H_{\beta,0}$ has a harmonic eigenspectrum: Eigenstates are $Q$-mode Fock states $\ket{m}_Q$, with eigenenergy $\delta_\beta m$. On top of this, we treat $V_\beta$ perturbatively. We note that one cannot neglect $V_\beta$: $H_{\beta,0}$ is quadratic, so the transmon looks like a linear harmonic cavity with a drive-renormalized frequency; the Kerr effect requires a nonlinear element and that comes only from $V_\beta$. We do the perturbation in the $Q$-modes picture. To that end, we write $V_\beta=V_3+V_4$ in terms of the $Q$-mode operators: $V_3 = -\alpha(V_{3,3} + V_{3,1})$ and $V_4 = -\tfrac{\alpha}{2}(V_{4,4}+V_{4,2}+V_{4,0})$, with (recalling $\vartheta=2\phi$)
\begin{align}
V_{3,3}&\equiv  \beta\big[c(1 + 3s^2) + s(2 + 3s^2)\big]\,\widehat Q^{\dagger 2}\widehat Q \\
&\quad +\upe^{\upi\vartheta}cs\beta(c + s)\,\widehat Q^{\dagger 3} +\textrm{h.c.}\nonumber\\
V_{3,1}&\equiv s\beta\big[1+3s(c+s)\big]\,\widehat Q^\dagger +\textrm{h.c.}\nonumber\\
V_{4,4}&\equiv {\left[\upe^{\upi2\vartheta}c^2 s^2\,\widehat Q^{\dagger 4}   + 2\upe^{\upi\vartheta}cs(1+2s^2)\,\widehat Q^{\dagger 3}\widehat Q+\textrm{h.c.}\right]}\nonumber\\
&\quad +(1 + 6 c^2s^2)\,\widehat Q^{\dagger 2}\widehat Q^2 \nonumber\\
V_{4,2}&\equiv {\left[\upe^{\upi\vartheta}cs\,(1 + 6s^2)\,\widehat Q^{\dagger 2} +\textrm{h.c.}\right]}+ 4s^2(2 + 3s^2)\,\widehat Q^\dagger \widehat Q \nonumber\\
V_{4,0}&\equiv  s^2(1 + 3s^2).\nonumber
\end{align}

We focus on the concrete case discussed in the main text, with $\overline m=1$, $\overline\sigma_a=+1$, and $\overline\sigma_b=-1$. We expect the dominant $q$s to be $q_{m'm}^\pm$s with nearby $m,m'=0,1$. Using the expressions for $V_\beta$ above, we can work out with perturbation theory the corrected eigenenergies: $\epsilon_m=\epsilon_m^{(0)}+\epsilon_m^{(1)}=\delta_\beta m + {}_Q\bra{m}V_4\ket m_Q$ (noting that $V_3$ is off-diagonal), with $
{}_Q\!\bra{m}V_4\ket m_Q=-\tfrac{\alpha}{2}{\left[(1+6c^2s^2)m(m-1)+4s^2(2+3s^2)m+s^2(1+3s^2)\right]}$.
This gives then, for the lowest two eigenstates, $\epsilon_0=-\tfrac{\alpha}{2}s^2(1+3s^2)$ and $\epsilon_1=\delta_\beta-\tfrac{3\alpha}{2}s^2(3+5s^2)$. The eigenstates of $H_\beta$, to first order in $V_\beta$, are
\begin{align}
\ket{\psi_{\beta;0}}&\equiv |0^{(0)}\rangle_Q+|0^{(1)}\rangle_Q\\
&= \ket 0_Q+\frac{\alpha}{\delta_\beta}\Bigl\{s\beta[1+3s(c+s)]\,|1\rangle_Q \nonumber\\
&\quad + \tfrac{1}{2\sqrt 2}\upe^{\upi\vartheta}cs(1+6s^2)|2\rangle_Q \nonumber\\
&\quad + \tfrac{\sqrt 6}{3}\upe^{\upi\vartheta}cs\beta(c + s)|3\rangle_Q 
+ \tfrac{\sqrt 6}{4}\upe^{\upi2\vartheta}c^2 s^2|4\rangle_Q\Bigr\}\,.\nonumber\\
\ket{\psi_{\beta;1}} &\equiv |1^{(0)}\rangle_Q+|1^{(1)}\rangle_Q\\
&=\ket 1_Q+ \frac{\alpha}{\delta_\beta}\Bigl\{-s\beta^*(1+3s^2 + 3cs)|0\rangle_Q \nonumber\\
&\quad + \sqrt 2\beta[c(1+6s^2) + 3s(1+2s^2)]|2\rangle_Q \nonumber\\
&\quad + \tfrac{\sqrt 6}{4}\upe^{\upi\vartheta}cs(3+10s^2)|3\rangle_Q + \tfrac{2\sqrt 6}{3}\upe^{\upi\vartheta}cs\beta(c + s)|4\rangle_Q \nonumber\\
&\quad + \tfrac{\sqrt{120}}{8}\upe^{\upi2\vartheta}c^2 s^2|5\rangle_Q\Bigr\}.\nonumber
\end{align}
Note that the eigenstates $\ket{\psi_m}$ of the original (undisplaced) $H_q^\textrm{RWA}$ are related via a displacement: $\ket{\psi_m}=\cD(\beta)\ket{\psi_{\beta;m}}$. From these, we can compute the dominant $q_{m'm}^{+}\equiv \bra{\psi_{m'}}q^{\dagger}\ket{\psi_m}=\bra{\psi_{\beta;m'}}(q^{\dagger}+\beta^*)\ket{\psi_{\beta;m}}$:
\begin{align}
q^+_{0,0} &= \beta^* {\left[1+ \tfrac{\alpha}{\delta_\beta}s(c+s)^2(c+2s)\right]},\nonumber\\
q^+_{1,1} &= \beta^* {\left[1+ \tfrac{\alpha}{\delta_\beta}(c+s)^2(2+3cs+6s^2)\right]},\nonumber\\
q^+_{1,0} &= c{\left[1+ \tfrac{\alpha}{2\delta_\beta}s^2(1+6s^2)\right]}, \nonumber\\
q^+_{0,1} &= \upe^{-\upi\vartheta}s{\left[1+ \tfrac{\alpha}{2\delta_\beta}c^2(1+6s^2)\right]};
\end{align}
the $q_{mm'}^-$s are available via complex conjugation.

For the chosen resonance $D^\textrm{res}$ with $\overline m=1$, $\overline\sigma_a=+1$ and $\overline\sigma_b=-1$, we have [see Eq.~\eqref{eq:KABres}]
\begin{align}
L &\simeq 
\frac{q_{1,0}^-(q_{0,0}^+-q_{1,1}^+)}{\delta_{ad}}-\frac{q_{1,0}^+(q_{0,0}^--q_{1,1}^-)}{\delta_{bd}}+O{\left(\frac{q^2}{\delta}\frac{D^\textrm{res}}{\delta}\right)},\nonumber
\end{align}
with $q$ representing the $q_{mm'}^\pm$s and $\delta$ being $\delta_{ad}$ or $\delta_{bd}$; the $O(\ldots)$ term vanishes exactly on resonance $D^\textrm{res}=0$. Putting in the $q_{mm'}^{\pm}$s from above into the exact on-resonance term yields the approximate formula for $K_{ab}$ [Eq.~\eqref{eq:KABamp}] in the main text.

We look momentarily at the strong-drive limit of $|\Omega_d/\delta_d|\gg 1$. $|\Omega_d|$ goes into $\cK$, so we have $|\cK|\gg |\frac{\alpha}{\delta}|$. Assuming $|\tfrac{\alpha}{\delta}|\gtrsim 1$, we then have $|\cK|\gg 1$. In this limit, Eq.~\eqref{eq:xi} tells us that $\xi$ is large so that $1+\xi\simeq\xi$, and $\xi\simeq\cK^{1/3}$. Then $|\beta|\simeq |\frac{\Omega_d}{\alpha}|^{1/3}$ assumed to be large. In this limit, $V_\beta$ can hardly be argued to be small, and the perturbative expressions above are poor approximations. We thus focus our attention on the intermediate drive-strength regime, which suffices for finding a reasonable working point for the \crot~operation.


\bigskip
\noindent\textbf{Exact numerics}\\[0.5ex]
Here, we describe the exact numerics used to optimize the working point. The goal is to extract the effective Kerr coefficients exactly, improving on the perturbative expressions of Eqs.~\eqref{eq:selfKerr} and \eqref{eq:crossKerr}. This is done by a numerical diagonalization of the full Hamiltonian $H^\textrm{RWA}$ [Eq.~\eqref{eq:RWA_H}] to obtain the exact eigenenergies of the two-cavity--driven-transmon system. For each eigenenergy value, the corresponding quantum numbers $\mu\equiv (m, N_a,N_b)$ are determined by finding the uncoupled basis state $\ket{\psi_m,N_a,N_b}$ with the maximal overlap, and requiring a one-to-one correspondence between coupled and uncoupled eigenstates. Whenever ambiguities arise, the quantum-number assignment is verified against the bare eigenvalue spectrum, confirming that the state's energy is consistent with the expected quantum-number ordering. The phase of each numerical eigenstate is fixed by requiring the overlap to be real and positive, so that the superposition built from these labeled states have the intended relative phase.

The eigenenergies can be expressed in powers of $N_a$ and $N_b$ (the cavity-$a$ and $b$ occupation numbers) \cite{Zhang2022} $\cE_m(N_a, N_b)=\sum_{p,q=0}^\infty k_{m;(p,q)} (N_a)^p (N_b)^q$; here, as usual, $m$ is the transmon quantum number. Since we are interested in an off-resonant drive for which the transmon remains in the ground state, only $m=0$ is relevant, and the Kerr coefficients are the coefficients of the quadratic terms in the expansion: $K_a=k_{0;(2,0)}$, $K_b=k_{0;(0,2)}$, and $K_{ab}=k_{0;(1,1)}$.

To find these Kerr coefficients, we make use of the numerical eigenenergies to determine a system of linear equations. For a chosen occupation-number pair $(N_a^{(i)}, N_b^{(i)})$, we define $y_i\equiv \cE_0^{(i)}-\cE_0-\delta_{ad}N_a^{(i)}-\delta_{bd}N_b^{(i)}$ where $\cE_0^{(i)}\equiv\cE_0(N_a^{(i)}, N_b^{(i)})$ and $\cE_0\equiv \cE_0(0,0)$ are values obtained from the exact diagonalization. Here, we have taken out the (comparatively) large and known $\delta_{ad}$ and $\delta_{bd}$, to focus on the remaining smaller corrections. Putting in the expansion of $\cE_0(N_a^{(i)},N_b^{(i)})$ in terms of the occupation numbers and $k$ coefficients, we find the relation: $y_i=(k_{0;(1,0)}-\delta_{ad})N_a^{(i)}+(k_{0;(0,1)}-\delta_{bd})N_b^{(i)}+K_a(N_a^{(i)})^2+K_b(N_b^{(i)})^2+K_{ab}N_a^{(i)}N_b^{(i)}+\ldots$, where the ellipsis denotes the higher-order $p,q$ terms. 

We choose a set of occupation-number pairs $S\equiv \{N_a^{(i)},N_b^{(i)}\}$, and recognizing that the $k$ coefficients are independent of the occupation numbers, we can reorganize the set of $y_i$s, for $i$ running over all pairs in $S$ into a linear system: $\bm y = \Lambda\bm k$, with $\bm y\equiv (y_i)_i$, $\Lambda$ is a matrix with entries $\Lambda_{i;(p,q)}\equiv (N_a^{(i)})^p(N_b^{(i)})^q$ ($i$ is the row index, and $(p,q)$ is the column (pair-)index for $\Lambda$, and $\bm k \equiv (k'_{0;(p,q)})_{(p,q)}$, where $k'_{0;(p,q)}\equiv k_{0;(p,q)}$ whenever $p+q>1$, $k'_{0;(1,0)}\equiv k_{0;(1,0)}-\delta_{ad}$, and $k'_{0;(0,1)}\equiv k_{0;(0,1)}-\delta_{bd}$. Here, $i$ is an index that runs over all occupation-number pairs in $S$. For a finite system, we consider only up to some maximal power of the occupation numbers, $1\leq p+q\leq d$, for some chosen truncation $d$ (see below). We thus have a linear system that can be used to determine $\bm k$, which contains the desired Kerr coefficients.  

Now, how do we choose $S$ and $d$? We need to be sure that we do not have an underdetermined system of equations, and furthermore, we want to choose the truncation such that the effect of the dropped higher-order terms is small. We must have that $|S|\geq d(d+3)/2(= |\bm k|)$, and the matrix $\Lambda$ must be full rank ($\textrm{rank}(\Lambda)=|\bm k|$), demanding that we must have that, for each $(p,q)$ within the truncation, there is at least one configuration with $N_a^{(i)}\geq p, N_b^{(i)}\geq q$; otherwise, the columns of $\Lambda$ are linearly dependent. This does not ensure $\Lambda$ is full rank, but we can check that explicitly for our choice of $S$ and $d$. To have lower influence from the dropped higher-order terms, we target low occupation numbers. We thus set $d=3$ and choose $N_a,N_b\in\{0,1,...,d\}, (N_a,N_b)\neq(0,0)$. The $\bm k$ are then solved by a least-squares approach, yielding (among other values) the desired Kerr coefficients.


\bigskip
\noindent\textbf{Self-Kerr cancellation}\\[0.5ex]
A driven side transmon coupled to each cavity can be introduced to cancel unwanted self-Kerr effects, using the scheme introduced in Ref.~\cite{Zhang2022}. The side transmon gives rise to an additional self-Kerr term, with sign and magnitude determined by the side-transmon parameters. The total self-Kerr on the cavity is a simple sum of the individual self-Kerr terms from the central and side transmons: Terms that involve both transmons jointly can only go through a single virtual level per transmon (at 4th-order); for such single-level transitions, the transmon anharmonicities---responsible for the nonlinearities---cannot be felt and hence no self-Kerr contributions arise. A similar reasoning explains why the cross-Kerr effect between the two cavities is unaffected by the side transmons. The side-transmon drive, however, modifies the higher-order nonlinearities, and one could choose a working point where some cancellation of those terms also occurs, for improved gate performance.

The frequency of the side-transmon drive can be set equal to that of the central transmon for single-frequency operation; the remaining parameters (bare side-transmon frequency, anharmonicity, and drive amplitude) provide adequate degrees of freedom for a suitable working point. Following Ref.~\cite{Zhang2022}, the side transmon is driven to induce a new resonance in the cavity self-Kerr spectrum (from side transmon only); one operates close or far away from that resonance depending on whether a large or small cancellation is called for. The formula for the self-Kerr coefficient [Eq.~\eqref{eq:selfKerr}] gives intuition on the sign of the added self-Kerr, so that a positive or negative cancellation can be applied on demand. In particular, focusing on the dominant terms close to the resonance, one can argue that $\textrm{sgn}(K_a)=-\textrm{sgn}(\sigma\delta_{ad}+\epsilon_{mm_1})$ when choosing $\sigma, m_1$ to match a single-photon resonance process, or $\textrm{sgn}(K_a)=\textrm{sgn}(2\sigma\delta_{ad}+\epsilon_{mm_2})$ when matching a two-photon resonance process. The resonance condition can be approached from $0^+$ or $0^-$ to achieve the desired self-Kerr term for cancellation.

\bigskip
\noindent\textbf{Parameter tolerance}\\[0.5ex]
As the driven system is only specified relative to $\omega_q$ by detunings in units of $\alpha$ and ratios $\frac{g}{\delta}, \frac{\Omega}{\delta}$, the Kerr spectrum is unchanged (in units of $\alpha$) regardless of the precise value of $\omega_{q}$, as long as the relative values of all components are maintained. Practical considerations may limit our ability to operate exactly at the optimized working point. For the non-drive parameters, we find a change in value of $\pm0.5\%\,\alpha$ or $\sim\!\!\pm$1MHz leads to $\sim\!\!5\%$ change in the Kerr values, while a $\pm 1$MHz uncertainty in the drive parameters leads to a greater change of $\sim\!\!20\%$ as we are operating near divergences. A mismatch in the self-Kerr cancellation drive is most detrimental, leading to a potential increase in the self-Kerr by an order of magnitude. Nevertheless, we note that even without perfect cancellation, the cavity self-Kerr can generally be suppressed to be much weaker than the amplified cross-Kerr; it thus contributes only insignificant distortion of cavity modes within the gate time of interest. The MHz uncertainty range considered here is also likely too severe; experimental characterization and controls can generally achieve a better precision in the order of kHz.

\bigskip
\noindent\textbf{Error correction example}\\[0.5ex]
We consider a memory logical qubit encoded with an RSB code. The EC circuit is shown in Fig.~\ref{fig:Numerics}(c), requiring two \crot~gates on three modes. For both \crot~gates, the input data mode is treated as mode $b$ in our \crot~implementation above; the ancillary and output data modes are mode $a$, and thus set to the same physical parameters needed to operate at the working point. The latter is not a requirement of the EC circuit design, but allows for single-frequency operation of both \crot s. 

We can apply the transmon drives as described in the main text to implement the \crot s, turning the drives on for the gate times $T_{\crot 1}^{\varphi=\pi}=\pi/K_{ab}$ and $T_{\crot 2}^{\varphi=\pi/4}=\pi/4K_{ab}$ as demanded by the \crot~angles in the EC circuit. For a specific choice of RSB code, however, we can further optimize performance by tuning the gate times and the self-Kerr cancellation drives to reduce mode distortions from the drive-induced self-Kerr and higher-order terms. This is done by comparing the evolution of a variety of code states under the full Hamiltonian ($H^\textrm{RWA}$) to what the evolution would have been had we assumed the Hamiltonian contained only linear, self-Kerr, and cross-Kerr terms (and no higher-order terms). From this, we extract self- and cross-Kerr coefficients $K_{ab}$ and $K_{a(b)}$ that closely fit the exact evolution across the states, but are slightly shifted from those predicted from code-state-agnostic exact numerics described earlier---these state-adjusted Kerr coefficients try to accommodate higher-order terms as if they were also coming from Kerr effects. These Kerr coefficients are then used to adjust the gate times and the self-Kerr cancellation drives. Note that the self-Kerr drives are assumed to be applied perfectly; the side transmons are not included in the full simulation to constrain the dimension of the physical system to a simulatable size. To reduce distortion on the teleported output state even further, the drive on the second \crot~is set at a weaker amplitude of $\left|\frac{\Omega_d}{\delta_d}\right|\approx0.95$, and the self-Kerr cancellation is modified accordingly.

To isolate the impact of an imperfect \crot~in the EC circuit, we consider only correctable loss and phase errors. Specifically, the input data mode is subjected to the noise channel $\cN\equiv \cA_\textrm{loss}\circ \cD_\textrm{ph}$. Here, $\cA_\textrm{loss}$ is the single-photon loss channel $\cA_\textrm{loss}(\cdot) \equiv A_0(\cdot)A_0^\dagger + A_1(\cdot)A_1^\dagger$, with $A_0 \equiv \upe^{-\frac{1}{2}\kappa T\widehat n_c}$ and $A_1=\eta\widehat c$, where $\widehat c$ is the annihilation operator on the relevant photon mode and $\widehat n_c$ is the associated number operator; $\eta$ is an operator-valued factor to enforce the trace-preservation nature of $\cA_\textrm{loss}$; $\kappa$ is the photon loss rate and $T$ is some appropriate length of time for the noise to act (see below). $\cD_\textrm{ph}$ is the dephasing channel $\cD_\textrm{ph}(\cdot) \equiv \int^\infty_{-\infty} \mathrm{d}\theta \,p_\textrm{ph}(\theta) \,\upe^{\upi\theta\widehat n_c}(\cdot)\upe^{-\upi\theta\widehat n_c}$, with $p_\textrm{ph}(\theta)\equiv \frac{1}{Z}w(\theta)\upe^{-\theta^2/(2\kappa_\textrm{ph}T)}$,  a zero-mean windowed Gaussian distribution; the phase error is restricted to the correctable range $-\pi/2N<\theta<\pi/2N$ by a weight function $w(\theta)$ that applies a cosine taper towards the tails of the distribution; $Z$ is the normalization constant. 

The noise acts on the input for time $T\equiv T_\textrm{wait} + T_\textrm{EC}$. $T_\textrm{EC}$ is the time taken to execute the EC circuit, and we assume it is dominated by the gate times of the two \crot s so that $T_\textrm{EC}=T_{\crot 1}^{\varphi=\pi} + T_{\crot 2}^{\varphi=\pi/4} \sim19\mu s$ for the parameters specified in the main text.
$T_\textrm{wait}$ is the wait time before the EC circuit is applied; for each pair of loss and dephasing rates, $T_\textrm{wait}$ is chosen by maximizing the fidelity ratio $R=F/F_\textrm{bare}$ up to a maximum time window $\leq 500\mu s$ ($\sim 25\,T_\textrm{EC}$,  comparable to the cavity coherence times as measured in Ref.~\cite{Copetudo2026}), recognizing that the time before EC is a choice that should be optimized for. The break-even region $R>1$ has $T_\textrm{wait}$ at the maximal $500\mu s$ for $10^{-3}\mu s^{-1}$ and $10^{-2}\mu s^{-1}$ loss rates, while $T_\textrm{wait}\sim 150\mu s$ for $10^{-1}\mu s^{-1}$ loss rate.

\section*{Acknowledgements}
We acknowledge funding support from the Singapore Ministry of Education (MOE-T2EP50222-0017). S.Q., A.C., and A.K. acknowledge the support of the Singapore National Quantum Scholarship Scheme (NQSS). AI tools (Claude Opus) were used to verify lengthy algebraic calculations and in identifying typographical errors during manuscript preparation. All equations and results were checked by the authors, and the authors assume full responsibility for the accuracy and integrity of the work.

\bibliography{CROT-implementation.bib}

\end{document}